\documentclass[10pt,prx,twocolumn,superscriptaddress,aps]{revtex4-2}

\usepackage{graphicx}
\usepackage{dcolumn}
\usepackage{bm}
\usepackage{amsmath}
\usepackage{amsfonts}
\usepackage{amssymb}
\usepackage{dcolumn}
\usepackage{epsfig}
\usepackage{graphicx}
\usepackage{latexsym}
\usepackage{setspace}
\usepackage{fancyhdr}
\usepackage{wrapfig}
\usepackage{textgreek}
\usepackage{ulem}
\usepackage{titlesec}
\usepackage{outlines}
\usepackage{multirow}
\usepackage{adjustbox}
\usepackage{enumitem}
\usepackage[official]{eurosym}
\usepackage{mdframed}
\usepackage{enumitem}
\usepackage{braket}
\usepackage{hhline}
\usepackage[title]{appendix}
\usepackage{comment}
\usepackage{titlesec}
\titlespacing{\section}{0pt}{1.5em}{1.5em}
\titlespacing{\subsection}{0pt}{1.5em}{1.5em}

\begin{document}



\title{A Versatile Chip-Scale Platform for High-Rate Entanglement Generation using an AlGaAs Microresonator Array}
\author{Yiming Pang}
\author{Joshua E. Castro}
\affiliation{\protect\hbox{Electrical and Computer Engineering Department, University of California, Santa Barbara, CA 93106, USA}}
\author{Trevor J. Steiner}
\affiliation{Materials Department, University of California, Santa Barbara, CA 93106, USA}
\author{Liao Duan}
\affiliation{Physics Department, University of California, Santa Barbara, CA 93106, USA}
\author{Noemi Tagliavacche}
\author{\\Massimo Borghi}
\affiliation{Department of Physics, University of Pavia, Italy}
\author{Lillian Thiel}
\author{Nicholas Lewis}
\affiliation{\protect\hbox{Electrical and Computer Engineering Department, University of California, Santa Barbara, CA 93106, USA}}
\author{\\John E. Bowers}
\affiliation{\protect\hbox{Electrical and Computer Engineering Department, University of California, Santa Barbara, CA 93106, USA}}
\affiliation{Materials Department, University of California, Santa Barbara, CA 93106, USA}
\author{Marco Liscidini}
\affiliation{Department of Physics, University of Pavia, Italy}
\author{Galan Moody}
\email{moody@ucsb.edu}
\affiliation{\protect\hbox{Electrical and Computer Engineering Department, University of California, Santa Barbara, CA 93106, USA}}


\begin{abstract}
Integrated photonic microresonators have become an essential resource for generating photonic qubits for quantum information processing, entanglement distribution and networking, and quantum communications. The pair generation rate is enhanced by reducing the microresonator radius, but this comes at the cost of increasing the frequency mode spacing and reducing the quantum information spectral density. Here, we circumvent this rate-density trade-off in an AlGaAs-on-insulator photonic device by multiplexing an array of 20 small-radius microresonators each producing a 650-GHz-spaced comb of time-energy entangled-photon pairs. The resonators can be independently tuned via integrated thermo-optic heaters, enabling control of the mode spacing from degeneracy up to a full free spectral range. We demonstrate simultaneous pumping of five resonators with up to $50$~GHz relative comb offsets, where each resonator produces pairs exhibiting time-energy entanglement visibilities up to 95$\%$, coincidence-to-accidental ratios exceeding 5,000, and an on-chip pair rate up to 2.6~GHz/mW$^2$ per comb line---more than 40 times improvement over prior work. As a demonstration, we generate frequency-bin qubits in a maximally entangled two-qubit Bell state with fidelity exceeding 87$\%$ (90$\%$ with background correction) and detected frequency-bin entanglement rates up to 7~kHz ($\sim 70$~MHz on-chip pair rate) using $\sim 250$~\textmu W pump power. Multiplexing small-radius microresonators combines the key capabilities required for programmable and dense photonic qubit encoding while retaining high pair-generation rates, heralded single-photon purity, and entanglement fidelity. 
\end{abstract}
\maketitle

\thispagestyle{plain}

\section{Introduction}
\indent Integrated photonic microresonators have become an indispensable component for photonic integrated circuits, driving breakthroughs in microcombs \cite{shu2022microcomb,pasquazi2018micro,gaeta2019photonic,kippenberg2011microresonator}, ultra-narrow linewidth lasers \cite{del2014self,lihachev2022platicon}, optical atomic clocks \cite{papp2014microresonator,newman2019architecture}, and quantum photonics \cite{moody20222022,guidry2022quantum}. The strength of integrated Kerr microcombs in particular lies in the phase coherence across the comb lines, leading to ultra-high stability and low parametric thresholds in highly nonlinear $\chi^{\left(3\right)}$ materials such as AlGaAs and silicon \cite{chang2020ultra}. When pumped below the threshold, frequency combs comprising few photons still retain two- or multi-photon phase coherence as verified through quantum interference experiments using interferometers \cite{franson1989bell,wakabayashi2015time}. The resulting quantum frequency comb, whereby two pump photons are annihilated to create signal and idler photons in adjacent sets of comb lines \cite{kues2019quantum}, has been directly used for encoding photonic qubits in a variety of degrees of freedom \cite{o2009photonic}, generating multi-photon entangled states \cite{reimer2016generation}, expanding the entanglement bandwidth \cite{lu2019chip}, and producing a comb of vacuum squeezed modes \cite{yang2021squeezed}.
\begin{figure*}[t!]
    \centering
    \includegraphics[width=\textwidth]{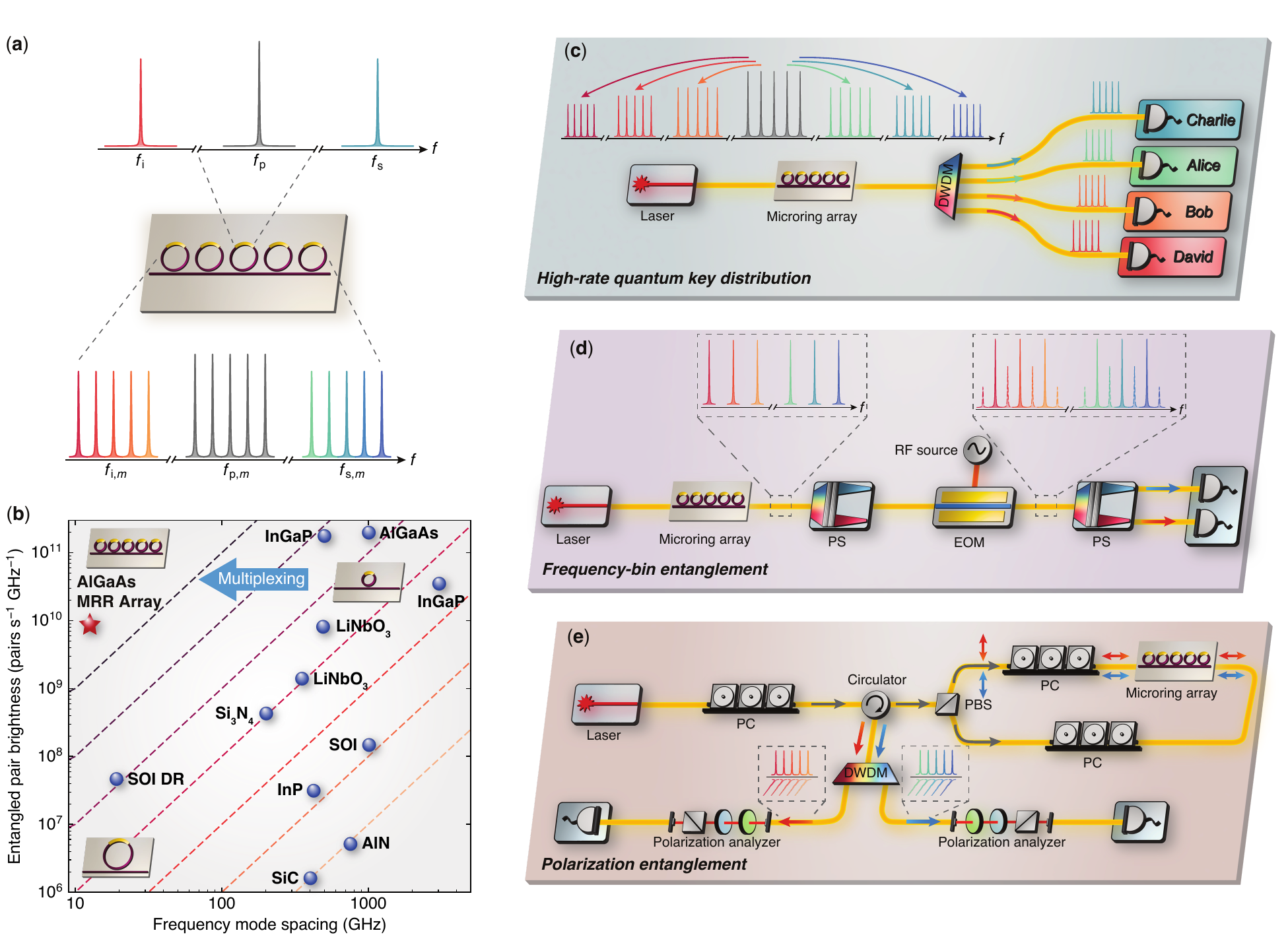}
    \vspace{-15pt}
    \caption{\small \label{Fig1} (\textbf{a}) Schematics showing the pump, signal, and idler frequency modes of a single microresonator (top) and groups of dense frequency modes from multiplexed microresonators (bottom). (\textbf{b}) Brightness and mode spacing for various entangled photon pair platforms \cite{steiner2021ultrabright, ma2017silicon, ramelow2015silicon, guo2017parametric, kumar2019entangled, zhao2022ingap, chopin2023ultra, ma2020ultrabright, rahmouni2024entangled, xu2022spectrally, clementi2023programmable}. For the AlGaAs microring resonator (MRR) array and the SOI dual-resonator (DR) designs, the brightness shown is for each resonator. (\textbf{c}) Schematics of experiment setups for high-rate quantum key distribution \cite{steiner2023continuous}, (\textbf{d}) frequency-bin entanglement \cite{imany201850}, and (\textbf{e}) multi-mode polarization entanglement \cite{miloshevsky2024cmos, takesue2008generation}, which can all benefit from dense spectral modes generated from a microresonator array device. (DWDM, dense wavelength division multiplexing filters; PS, pulse shaper; EOM, electro-optic modulator; PC, polarization controller; PBS, polarization beam splitter.)}
    \vspace{-10pt}
\end{figure*}
\\
\indent The choice of the microresonator material and design can be tailored to meet the requirements of different quantum information applications, most of which demand quantum photonic state generation with high rate, indistinguishability, fidelity, and heralded single-photon purity. For telecommunications wavelengths, over a decade ago it was revealed that the resonant enhancement in silicon-on-insulator (SOI) microcavities significantly improved the pair generation rate (PGR) compared to waveguide structures, which scales as $R_{\text{PG}} \propto n_2^2 Q^3 V^{-2} P^2$ with $n_2$ the Kerr nonlinear index, $Q$ the average quality factor of the pump, signal, and idler frequency modes, $V$ the resonator volume, and $P$ the on-chip pump power \cite{clemmen2009continuous,engin2013photon,azzini2012ultra}. Because the PGR improves with nonlinearity and quality factor (inversely proportional to the cavity round-trip loss), pair generation with Kerr microcombs has been examined in a variety of material platforms, including InGaP \cite{chopin2023ultra,kumar2019entangled}, Si$_3$N$_4$ \cite{ramelow2015silicon}, LiNbO$_3$ \cite{xu2022spectrally,zhao2020high}, SiC \cite{rahmouni2024entangled}, GaN \cite{zeng2024quantum}, and AlGaAs \cite{baboux2023nonlinear}. The high $\chi^{\left(2\right)}$ and $\chi^{\left(3\right)}$ optical nonlinearities of III-V materials make them particularly attractive, with recent demonstrations of pair generation rates exceeding 20~GHz/mW$^2$ for microresonators and photonic crystal cavities with AlGaAs \cite{steiner2021ultrabright} and InGaP \cite{chopin2023ultra}, respectively, and time-bin entanglement-based quantum key distribution of nearly $10,000$ entangled bits~s$^{-1}$ using deployed optical fibers \cite{steiner2023continuous}. \\

\indent High pair-generation rates come with a trade-off, however; confining the optical modes into small-volume ($V \propto 2\pi R$, where $R$ is the resonator radius) resonators quadratically improves the pair-generation rate, but at the expense of a large cavity free-spectral range (FSR). As shown in Fig.~\ref{Fig1}, the highest PGRs have been demonstrated with $\sim$1~THz combs, which is significantly larger than the 100 GHz telecom C-band ITU grid. This trade-off between the rate and mode spacing reduces the quantum information spectral density and prevents the full use of commercial off-the-shelf dense wavelength division multiplexing (DWDM) components. Additionally, the large mode spacing is incompatible with frequency-bin photonic qubit encoding \cite{imany201850}, whereby the frequency modes of different comb lines are defined as the qubit states that can be mixed and interfered using electro-optic modulators (EOM) and pulse shapers (PS) \cite{lu2023frequency}.
\begin{figure*}[t!]
    \centering
    \includegraphics[width=\textwidth]{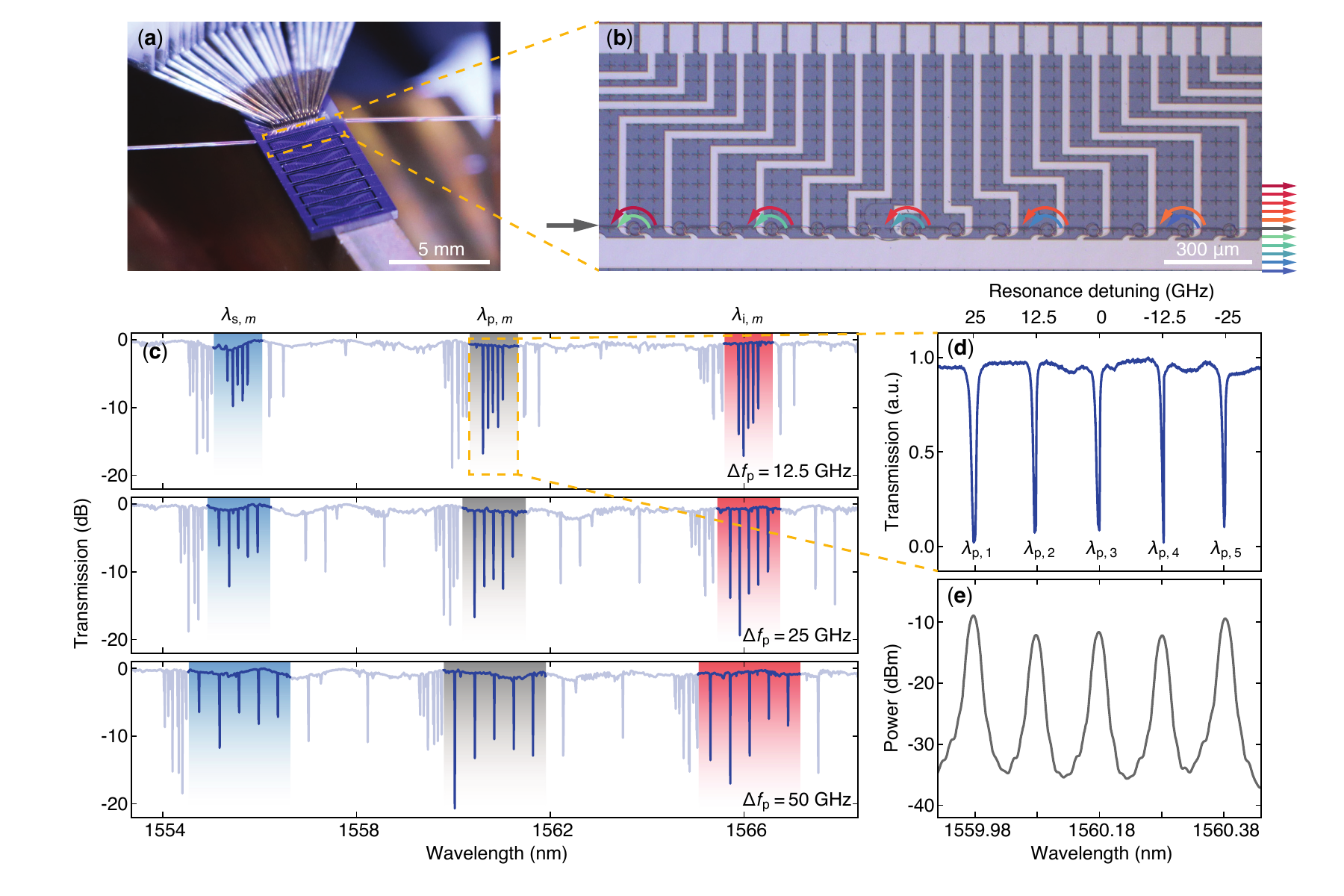}
    \vspace{-15pt}
    \caption{\small \label{Fig2} (\textbf{a}) Image of an AlGaAsOI chip with 14 sets of microring array devices with fiber input and output coupled to one array. (\textbf{b}) Micrograph of the device showing an array of 20 microresonators coupled to a single bus waveguide. Each microresonator has metal heaters used to independently thermo-optically tune resonances with sub-GHz precision. With the resonances slightly detuned, spectral modes from all resonators can be multiplexed into a single bus waveguide for on/off-chip entanglement distribution. (\textbf{c}) Transmission spectra of the device showing five detuned resonators with pump ($\lambda_{\text{p},m}$), signal ($\lambda_{\text{s},m}$), and idler ($\lambda_{\text{i},m}$) modes with $\Delta f_{\text{p}}$~=~12.5~GHz (top), 25~GHz (center), and 50~GHz (bottom). (\textbf{d}) A zoom-in of the pump resonances with 12.5~GHz detuning and (\textbf{e}) the power spectrum of a 12.5~GHz repetition rate electro-optic (EO) modulated laser showing excellent alignment to drive the pump resonances.}
    \vspace{-10pt}
\end{figure*}

\thispagestyle{plain}

\indent Here, we demonstrate a photonic device that enables dense wavelength division multiplexing of microcombs using an array of up to 20 integrated AlGaAs-on-insulator microresonators coupled to a single bus waveguide. This concept is motivated by a recent theoretical study \cite{liscidini2019scalable} and experimental demonstrations of silicon multi-resonator devices \cite{andrea2022silicon,borghi2023reconfigurable} that enabled programmable two-qubit Bell-state generation and frequency encoded quantum key distribution \cite{clementi2023programmable,tagliavacche2024}. By using AlGaAs instead of silicon, we enhance the pair generation rate by $> 40$ times from prior dual-resonator designs and $> 2,500$ over single resonator designs for sub-20~GHz frequency mode spacing. Each resonator in the array exhibits high two-photon interference visibility up to 95$\%$, single-photon purity up to $99\%$, coincidence-to-accidental ratio up to $5,000$, and on-chip rates exceeding 30 million pairs per second with less than 100~\textmu W of on-chip pump power. The resonators can be independently tuned with their respective thermo-optic heaters, allowing the individual microcombs to be interleaved with offsets spanning degeneracy up to the full 650~GHz FSR. With high-rate and high-quality pair generation with tunable microcomb spacing, new opportunities arise in tunable indistinguishability \cite{alexander2024manufacturable}, high-rate quantum key distribution (Fig.~\ref{Fig1}(\textbf{c})), frequency bin entanglement (Fig.~\ref{Fig1}(\textbf{d})), and polarization entanglement (Fig. \ref{Fig1}(\textbf{e})). As an illustrative example, we tune two resonators to a 36~GHz comb offset and simultaneously pump them to create frequency encoded qubits, and with an off-chip modulator and filters, we show frequency bin entanglement with fidelity up to 87$\%$ (90$\%$ with background subtraction) and up to 7~kHz detected entangled pair rates, exceeding previous demonstrations despite the high insertion loss of our off-chip modulator and filter components.
\begin{figure*}[t!]
    \centering
    \includegraphics[width=1\textwidth]{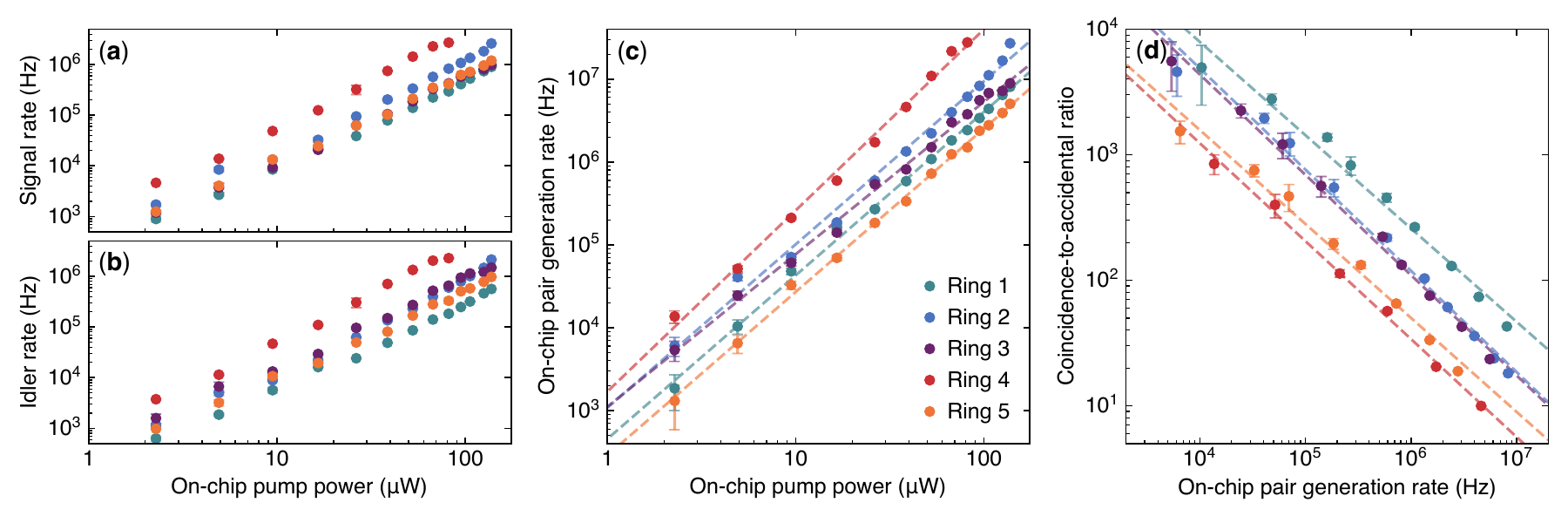}
    \vspace{-15pt}
    \caption{\small \label{Fig3} (\textbf{a}) Detected signal and (\textbf{b}) idler count rates versus on-chip pump power. Measurement uncertainty is determined by taking the standard deviation of count rates over $>60$~seconds integration time. (\textbf{c}) On-chip pair generation rate versus on-chip pump power determined after accounting for loss in the measurement setup. The highest internal pair generation efficiency is measured to be 2.6~GHz/mW$^2$ (ring 4). (\textbf{d}) Coincidence-to-accidental ratio (CAR) versus on-chip pair generation rate. Dashed lines are fits of the experimental data. Maximum CAR $>5,000$ is measured for $10^4$ pairs s$^{-1}$ on chip.}
    \vspace{-10pt}
\end{figure*}
\vspace{-7pt}

\thispagestyle{plain}

\section{Device Design and Frequency Mode Tuning} 
\indent Figure \ref{Fig2}(\textbf{a}) shows an example of our AlGaAsOI chip with 14 sets of microresonator array devices that is fabricated with a 400~nm thick Al$_{0.2}$Ga$_{0.8}$As photonic layer bonded on a 3~\textmu m SiO$_{2}$-on-silicon wafer. The detailed bonding, substrate removal, lithography, and etching processes are presented previously in \cite{xie2020ultrahigh, steiner2021ultrabright}. As shown in  Fig.~\ref{Fig2}(\textbf{b}), the device consists of 20 microresonators coupled to a single bus waveguide using pulley couplers at the ring-bus coupling region. Each resonator is nominally identical with an FSR of 650~GHz, an average loaded quality factor of $Q_{\mathrm{L}} \sim 3.5\times10^5$, and waveguide width of 800~nm to achieve zero dispersion for quasi-phase-matching. Individual metal heaters are deposited above the 1-\textmu m-thick top oxide cladding to tune each resonator. At the two ends of the bus waveguide, an adiabatic edge coupler with inverse taper design is used to couple light on and off chip to match the mode profile of our lensed polarization-maintaining optical fibers. The device presented in this work has coupling loss of 6.5~dB/facet since the facets were defined only via blade dicing, which can be mitigated by polishing or etching the facet and using anti-reflection coatings to reduce the loss to approximately $2-3$~dB/facet as previously demonstrated in \cite{castro2022expanding}. 
\\
\indent A probe card powered by a multichannel DC current source is placed in contact with the metal pads to provide electrical access to the device. By sweeping current on all channels, resonances of each microresonator can be identified and then individually tuned with sub-GHz precision and arbitrary spacing between the resonances. The three panels in Fig.~\ref{Fig2}(\textbf{c}) show transmission spectra from the device where we thermally tune resonances of five different resonators (the other 15 resonators are not used for this demonstration) into equal frequency offset with $\Delta f_{\text{p}}=12.5$~GHz (top), 25~GHz (center), and 50~GHz (bottom), respectively. Groups of densely packed signal ($\lambda_{\text{s}, m}$), pump ($\lambda_{\text{p}, m}$) and idler ($\lambda_{\text{i}, m}$) frequency modes are highlighted. As demonstrated in Fig.~\ref{Fig2}(\textbf{d, e}), the 12.5~GHz detuned pump resonances show excellent alignment with the power spectrum of a 12.5~GHz spaced electro-optic (EO) modulated continuous-wave pump laser. The microresonators can be simultaneously pumped by the EO comb to generate spectrally dense quantum microcombs that are multiplexed in the bus waveguide and can be routed off-chip for entanglement distribution.
\section{Characterization of Pair Sources}
\subsection{Pair Generation Rate and CAR}
\indent To characterize the pair generation efficiency of the device, resonances of five microresonators are detuned with $\Delta f_{\text{p}}$~=~12.5~GHz (Fig.~\ref{Fig2}(\textbf{c}), top) such that all sets of signal and idler photons can be isolated with a pair of 100~GHz DWDM filters. Five microresonators are individually pumped with a tunable continuous-wave laser followed by a series of tunable bandpass filters before the chip to suppress amplified spontaneous emission sideband from the laser. Time-correlated photon pairs are generated and filtered off-chip using transmission and reflection ports of the DWDM filters (ITU channel~28 for signal photons near 1554.94~nm and channel~15 for idler photons near 1565.5~nm). Photons are sent to a pair of superconducting nanowire single-photon detectors (SNSPDs) with 85\% detection efficiencies. Photon counting events are time-tagged with a time-correlated single-photon counter (TCSPC). The cumulative losses from bus waveguide to the detectors are 13.5~dB and 12.5~dB for signal and idler channels, respectively. \\

\thispagestyle{plain}

\indent Detected signal and idler count rates at on-chip pump power levels in the range of 1-100~\textmu W are shown in Fig.~\ref{Fig3}(\textbf{a, b}). Background counts from residual pump and room light leakage are measured separately (tuning the laser wavelength off-resonance) and subtracted from detected raw counts. The maximum detected singles rate is $\sim3$~MHz since our SNSPDs would latch beyond this level. An exemplary coincidence count histogram is shown in Appendix \ref{Appendix A}, where the detected coincidences $N_{\text{cc}}$ are determined by fitting the data with a Gaussian function and integrating over its full width at half maximum (FWHM). The uncertainty in $N_{\text{cc}}$ is estimated by assuming that the photon-counting process satisfies Poissonian statistics. The on-chip pair generation rate shown in Fig.~\ref{Fig3}(\textbf{c}) is calculated as $R_{\text{PG}}=N_{\text{cc}}/\left(\Delta t \eta_{\text{s}}\eta_{\text{i}}\right)$ where $\eta_{\text{s}}$ and $\eta_{\text{i}}$ are detection efficiencies for the signal and idler channels, respectively, and $\Delta t$ is the integration time of the measurement (varying for each pump power but $>60$~seconds for all data). We extract an average internal pair generation efficiency of 1~GHz/mW$^{2}$ for all five resonators, and the highest efficiency of 2.6~GHz/mW$^{2}$ (ring 4) is on par with state-of-the-art Kerr microresonator entangled-pair sources \cite{steiner2021ultrabright, chopin2023ultra}. Coincidence-to-accidental ratio (CAR) is calculated using $\text{CAR}=(N_{\text{cc}}-N_{\text{acc}})/N_{\text{acc}}$ with $N_{\text{acc}}$ being the accidental coincidence counts integrated over the same integration window (FWHM of coincidence peak) but away form the coincidence peak at $\tau=0$~ns (Appendix \ref{Appendix A}). CAR higher than 5,000 is measured at $R_{\text{PG}}\sim 10$~kHz, and the expected inverse relation between CAR and PGR is observed in Fig.~\ref{Fig3}(\textbf{d}). The measured $N_{\text{cc}}$ and $N_{\text{acc}}$ are fitted with pump power using a polynomial fit, and the results suggest $N_{\text{cc}}\propto P^{2}$ and $N_{\text{acc}}\propto P^{4}$, which indicate that the sources operate in the low nonlinear noise regime with negligible contributions from, e.g. spontaneous Raman scattering in fiber and the waveguide \cite{guo2017high}. 
\begin{figure*}[t!]
    \centering
    \includegraphics[width=\textwidth]{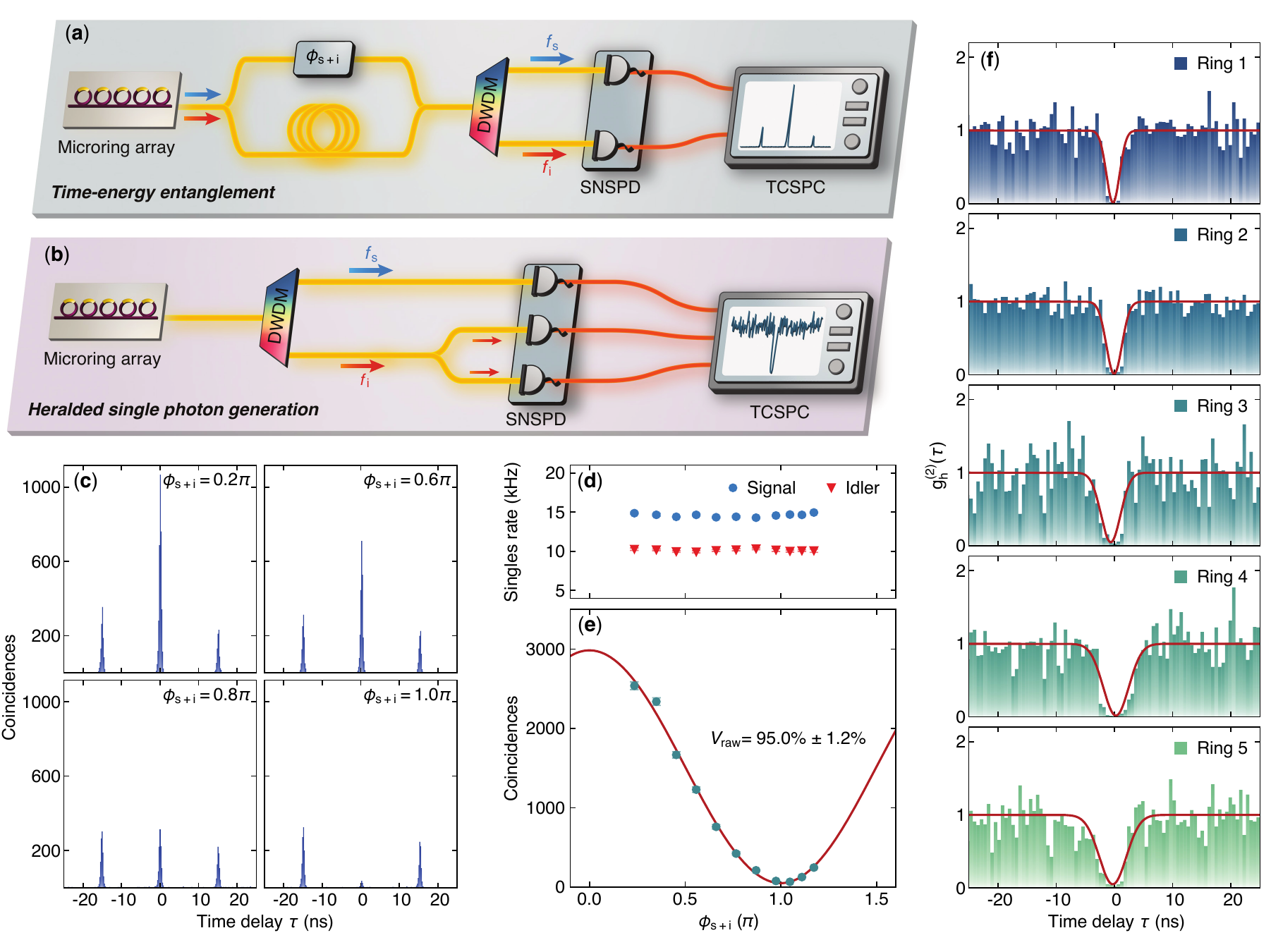}
    \vspace{-15pt}
    \caption{\small \label{Fig4} Schematic illustrations of the setup for (\textbf{a}) time-bin two-photon interference experiment using a fiber-based folded Franson interferometer, and (\textbf{b}) heralded single-photon generation by using the signal photon as herald and splitting the idler photon with a 50:50 fiber beam splitter. (SNSPD, superconducting nanowire single-photon detector; TCSPC, time-correlated single-photon counter.) (\textbf{c}) Coincidence histograms of time-bin entanglement experiment when the two-photon states have a phase difference of $\phi_{\text{s}+\text{i}}= 0.2\pi, 0.6\pi, 0.8\pi$ and $1.0\pi$, respectively. (\textbf{d}) Signal and idler count rates and (\textbf{e}) coincidences of the central peak versus phase. At 0.6~MHz on-chip rate (ring 1), the raw visibility is 95\%~$\pm$~1.2\%. (\textbf{f}) Heralded second-order correlation function $g_{\text{h}}^{(2)}(\tau)$ measured at $\sim 0.6$~MHz on-chip rate for five microresonators. $g_{\text{h}}^{(2)}(0)$ as low as 0.010~$\pm$~0.002 (ring 1) is measured corresponding to heralded single-photon purity of $99\%$.
    }
    \vspace{-10pt}
\end{figure*}

\thispagestyle{plain}

\subsection{Time-Bin Visibility and Heralded Purity}
\begin{figure*}[t!]
    \centering
    \includegraphics[width=\textwidth]{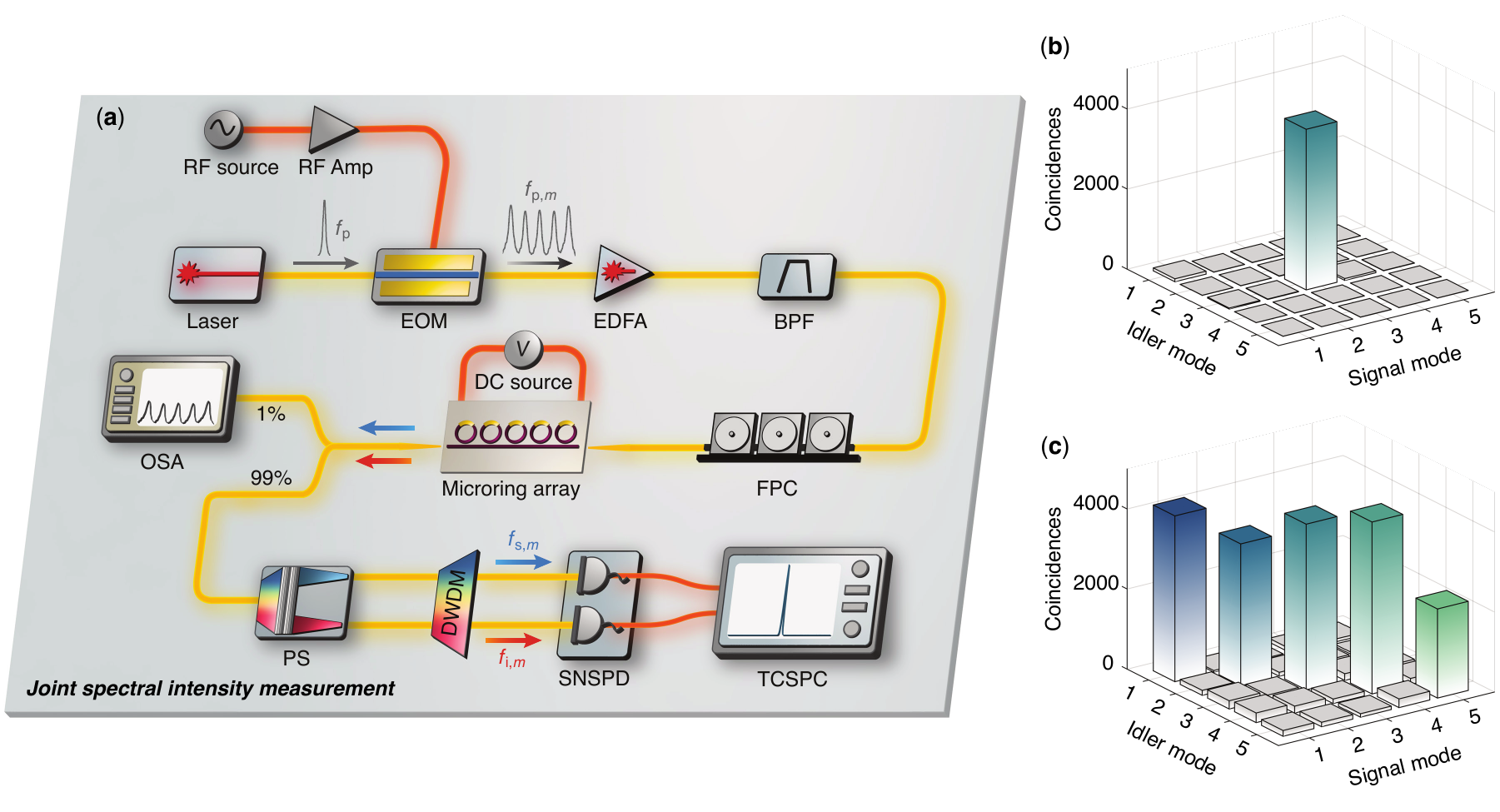}
    \vspace{-15pt}
    \caption{\small \label{Fig5} (\textbf{a}) A schematic illustration of the setup for joint spectral intensity (JSI) measurements. An electro-optic frequency comb simultaneously pump all five micorings, and a pulse shaper is programmed to scan both signal and idler filters with step size of 12.5~GHz. (RF Amp, RF amplifier; EDFA, erbium-doped fiber amplifier; BPF, tunable bandpass filters; FPC, fiber polarization controller; OSA, optical spectrum analyzer.) (\textbf{b}) JSI when the resonance of only one microring (ring 3) is aligned to the EO comb. (\textbf{c}) JSI when resonances of five microrings are aligned to the EO comb, generating pairs in all five frequency modes along the diagonal of the $5 \times 5$ frequency space. }
    \vspace{-10pt}
\end{figure*}
\indent Due to the spontaneous four-wave mixing process in Kerr microresonators, the generated photon pairs can exhibit time-energy correlations that we utilize to create time-bin entangled photonic qubits. This is achieved by sending the pairs into a Franson-like interferometer where the interference between the two-photon wavefunction can generate entanglement \cite{brendel1999pulsed, franson1989bell}. Figure~\ref{Fig4}(\textbf{a}) shows the experiment setup used for time-bin entanglement, where the generated pairs are sent into a fiber-based folded Franson interferometer with a long path and a short path (time delay $\Delta \tau \approx 16$~ns) that can be expressed in the single-photon states as $\ket{\mathrm{L}}$ and $\ket{\mathrm{S}}$, respectively. A fiber phase shifter on the short path is used to sweep the relative phase between the two arms of the interferometer, which is actively stabilized by a servo control loop (not shown) that monitors the residual pump power and feeds the error signal to the phase shifter. When time-correlated pairs enter the interferometer, signal and idler photons can either take different paths resulting in distinguishable arrival time delay $\Delta \tau$ on the detectors, namely the two separable states $\ket{\mathrm{LS}}$ and $\ket{\mathrm{SL}}$ where index $i$($j$) in $\ket{ij}$ represents the path that signal (idler) photon travels. Alternatively, both signal and idler photons can take the same path and be detected at the same time, which are indistinguishable at the detectors, thereby creating an entangled state:
\begin{equation}
    \ket{\psi} = \frac{\ket{\mathrm{LL}}+e^{i\phi_{\mathrm{s}+\mathrm{i}}}\ket{\mathrm{SS}}}{\sqrt{2}}
\end{equation}
with $\phi_{\mathrm{s}+\mathrm{i}}$ being the cumulative phase applied to photon pairs passing through the phase shifter. 
\\
\indent Coincidence histograms in Fig.~\ref{Fig4}(\textbf{c}) provide direct visualization of the interferometer output: the phase-independent side peaks at $\tau = -16$~ns and $\tau = 16$~ns correspond to photon pairs in $\ket{\mathrm{LS}}$ and $\ket{\mathrm{SL}}$ states, respectively; and the central peak at $\tau = 0$~ns signifies the entangled state $\ket{\psi}$ where coincidence counts decrease as $\phi_{\mathrm{s}+\mathrm{i}}$ increases from $0.2\pi$ to $1.0\pi$---a consequence of destructive interference between the indistinguishable $\ket{\mathrm{LL}}$ and $\ket{\mathrm{SS}}$ states. An exemplary interferogram in Fig.~\ref{Fig4}(\textbf{e}) shows coincidence counts versus phase with singles count rates monitored during the measurement (Fig.~\ref{Fig4}(\textbf{d})). At 0.6 MHz on-chip PGR (ring 1), the raw interference visibility without background subtraction $V_{\mathrm{raw}}$ is 95.0\%~$\pm$~1.2\%, strongly violating the Clauser-Horne-Shimony-Holt (CHSH) inequality of $V > 70.7$\% by more than 20 standard deviations \cite{clauser1969proposed, kwiat1993high}. All five resonators generate time-bin entangled photon pairs at $\sim 0.6$~MHz internal rate with $V_{\mathrm{raw}} > 89$\%, as summarized in Table~\ref{Table 1}. 
\\
\indent The microresonator array device can also be used as multiplexed heralded single-photon sources. To assess the single-photon nature of pairs produced from each resonator, a Hanbury-Brown-Twiss experiment illustrated in Fig.~\ref{Fig4}(\textbf{b}) is performed by splitting the idler photon to two detectors with a 50:50 fiber beam splitter. The three-fold coincidence rate $R_{\text{s12}}(\tau)$ and the two-fold coincidence rates $R_{\text{s1}}(\tau)$ and $R_{\text{s2}}(\tau)$ are measured while monitoring the signal count rate $R_{\text{s}}$, and the heralded second-order correlation function $g_{\text{h}}^{(2)}(\tau)$ is expressed as \cite{signorini2020chip}:
\begin{equation}
    g_{\text{h}}^{(2)}(\tau) = \frac{R_{\text{s12}}(\tau) R_{\text{s}}(0)}{R_{\text{s1}}(0) R_{\text{s2}}(\tau)}
\end{equation}
\indent Figure~\ref{Fig4}(\textbf{f}) shows $g_{\text{h}}^{(2)}(\tau)$ measured at $\sim 0.6$~MHz on-chip rate for five microresonators with an integration time of 2 hours. $g_{\text{h}}^{(2)}(0)$ dips below 0.5 at $\tau = 0$ ns confirming photon anti-bunching, and the lowest $g_{\text{h}}^{(2)}(0)$ measured is 0.010~$\pm$~0.002 for ring 1, which translates to a heralded single-photon purity of 99\%. For all five resonators we obtain heralded single-photon purities greater than 97.5\% (Table \ref{Table 1}). 

\thispagestyle{plain}

\section{Frequency Bin Qubits and Entanglement}
\subsection{Joint Spectral Intensity of Microresonator Array}
\indent To examine the spectral multiplexing and any potential cross-talk or leakage between the resonator modes, we implement the setup in Fig.~\ref{Fig5}(\textbf{a}) to measure the joint spectral intensity (JSI) of the microresonator array. Pump light ($f_{\text{p}}$) enters the input of a commercial phase modulator driven by an amplified 12.5~GHz RF signal, and phase-modulated sidebands with 12.5~GHz optical frequency spacing ($f_{\text{p},m}$) form an electro-optic (EO) frequency comb. The pump EO comb is optically amplified and filtered and then used to pump all five microresonators with their resonances detuned by 12.5~GHz. An optical spectrum analyzer (OSA) is used to monitor the EO comb profile of the residual pump, and a programmable pulse shaper acts as a 10~GHz bandwidth tunable filter to selectively route signal ($f_{\text{s},m}$) and idler ($f_{\text{i},m}$) photons produced from each microresonator to the SNSPDs. To test the pulse shaper filter extinction ratio between adjacent frequency modes ($\sim 20$~dB per channel verified by separate measurements), we first aligned the resonances of one resonator (ring 3) to the EO comb, then performed a $5\times5$ scan in both signal and idler frequency modes by sweeping the pulse shaper filter central frequency in steps of 12.5~GHz. As expected, the JSI in Fig.~\ref{Fig5}(\textbf{b}) shows that coincidences appear only at the signal mode $f_{\text{s},3}$ and idler mode $f_{\text{i},3}$, with negligible background coincidences at other modes. 
\\
\indent Next, we align resonances of all five resonators to the pump comb, then repeat the JSI measurement to collect pairs generated from each resonator. The RF power driving the EOM and bandpass filter center wavelength are optimized to adjust EO comb power per line in order to compensate for differences in pair generation rates of the resonators. Figure~\ref{Fig5}(\textbf{c}) shows the JSI with five peaks along the diagonal of the $5 \times 5$ frequency space (12.5~GHz step size), with each diagonal peak representing correlated photon pairs generated from one of the five resonators. The background at off-diagonal modes is mainly due to accidental coincidences caused by signal and idler photons that originate from two different resonators and thus exhibit no temporal correlation. In this demonstration, the spectral density of the quantum microcombs is limited by our pulse shaper bandwidth; however, with recent progress in on-chip microring-resonator-based photonic pulse shapers \cite{cohen2024silicon}, even smaller mode spacing can be attained to fully explore the potential of multiplexing microresonators with higher $Q$ (narrower linewidth). 
\begin{figure*}[hbt]
    \centering
    \includegraphics[width=\textwidth]{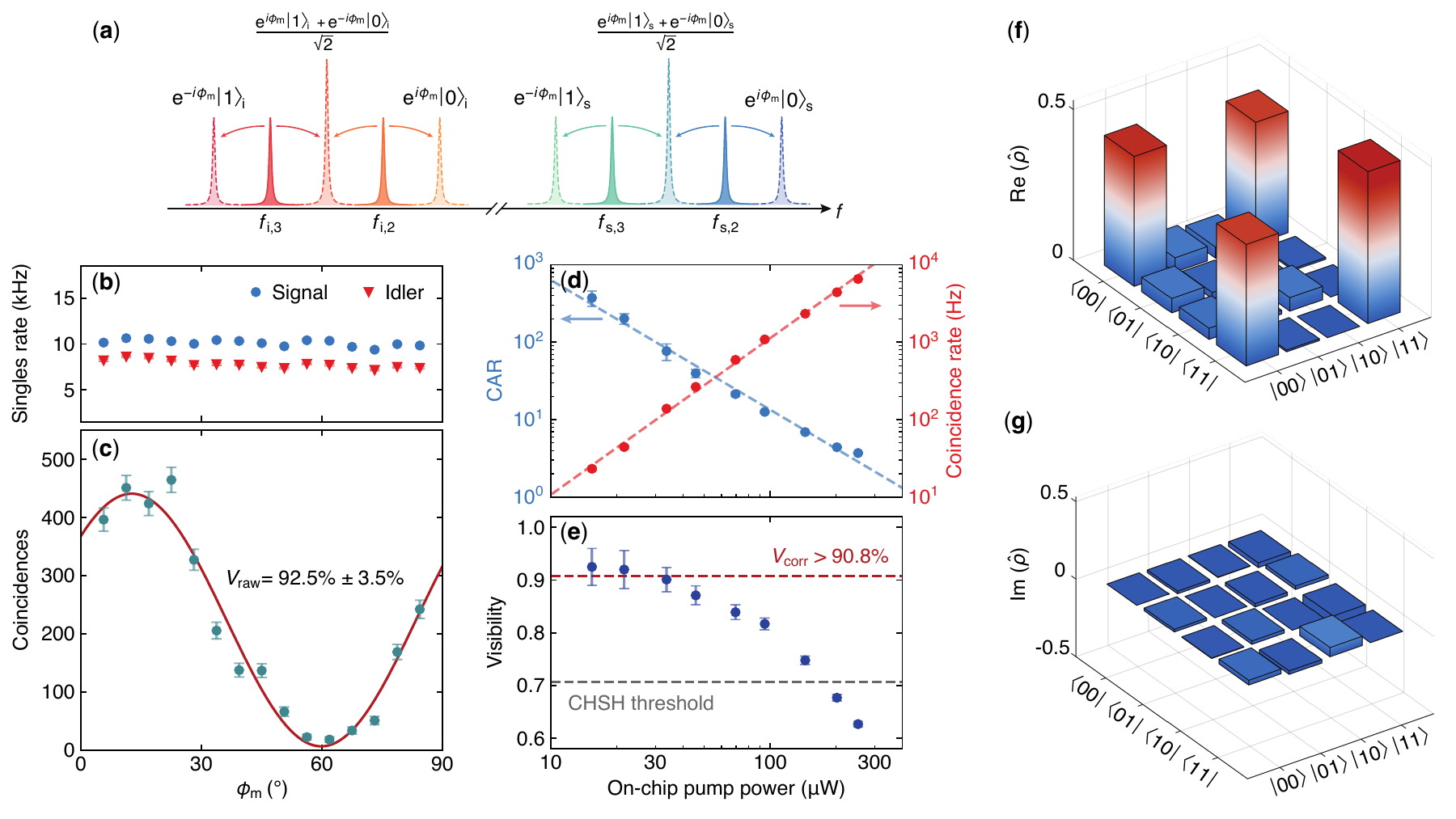}
    \vspace{-15pt}
    \caption{\small \label{Fig6} (\textbf{a}) Schematic of frequency mixing applied to the signal (blue) and idler (red) photons via phase modulation. The solid peaks represent input states: $\ket{0}_{\text{s,i}}$ from ring 2, and  $\ket{1}_{\text{s,i}}$ from ring 3. Dashed peaks are results of the frequency mixing: a central frequency bin is created by mixing photons from both resonators while two outermost bins are produced by converting photons from individual resonators up/down in frequency. Since photons converted up or down in frequency acquire a phase shift of $\pm \phi_{\text{m}}$ from the modulation phase $\phi_{\text{m}}$ of the mixing RF signal, a superposition state is formed at the central bin. (\textbf{b}) Signal and idler count rates, and (\textbf{c}) coincidences of the mixed frequency bin for varying modulation phases. At a CAR of 375, the raw visibility is 92.5\%~$\pm$~3.5\%. (\textbf{d}) CAR and detected coincidence rate at the mixed bin, and  (\textbf{e}) raw entanglement visibility $V_{\text{raw}}$ for different on-chip pump power. Grey dashed line indicates CHSH threshold of 70.7\%. Red dashed line shows visibility after background accidentals correction $V_{\text{corr}}$ remains above 90.8\% even at detected coincidence rate up to 7~kHz. (\textbf{f}) Real and (\textbf{g}) imaginary parts of the reconstructed density matrix $\hat{\rho}$ with fidelity $> 87\%$ for the maximally entangled Bell state.
    }
    \vspace{-10pt}
\end{figure*}

\thispagestyle{plain}

\subsection{Frequency-Bin Entanglement}
\indent To demonstrate the device capabilities here, we focus on frequency-bin qubit entanglement generation (Fig.~\ref{Fig6}). Two resonators are tuned to 36~GHz comb offset and are pumped with the $\pm 1$ sidebands of an 18 GHz EO comb. Ring 2 and ring 3 are chosen for this experiment due to their similar pair generation efficiency (1.01~GHz/mW$^2$ and 0.77~GHz/mW$^2$), FWHM of resonances (616~MHz and 641~MHz), and free spectral range (650.67~GHz and 650.87~GHz). Tunable bandpass filters at the chip input are tuned to adjust the power of each pump comb line to ensure the two resonators generate pairs at the same rate. The generated pairs are sent to a second EOM driven by the same 18~GHz RF source, yet with a coherent and independent modulation phase $\phi_{\text{m}}$ addressed by an RF phase shifter relative to the RF signal on the first EOM producing the pump EO comb (Appendix \ref{Appendix B}). The modulation depth of the RF signal is optimized such that optical power transferred to the first-order sidebands is maximized while higher-order harmonics are minimized. 
\\
\indent When an EOM with input optical carrier frequency $f_{0}$ is driven by a sinusoidal signal of $V(t) = V_{0} \sin(2\pi f_{\text{m}}t+\phi_{\text{m}})$, the $n$th sideband of the generated harmonics has a frequency of $f_{n} = f_{0} + n f_{\text{m}}$ and phase of $\phi_{n} = n(\pi/2 + \phi_{\text{m}})$ relative to the optical carrier \cite{capmany2010quantum}. Hence, when a nonzero RF modulation phase $\phi_{\text{m}}$ is applied, the $+1$ sideband $f_{1}$ acquires an additional optical phase $\phi_{\text{m}}$ while the $-1$ sideband $f_{-1}$ acquires $-\phi_{\text{m}}$, effectively creating a phase difference of $2\phi_{\text{m}}$ between the two. A schematic illustration of the modulated comb lines is shown in Fig.~\ref{Fig6}(\textbf{a}), whereby the frequency mixing is applied to the signal (blue) and idler (red) photons at the second EOM. We denote the input states represented in solid peaks as $\ket{0}_{\text{s,i}}$ for signal ($f_{\text{s},2}$) and idler ($f_{\text{i},2}$) photons generated from ring 2, and $\ket{1}_{\text{s,i}}$ for signal ($f_{\text{s},3}$) and idler ($f_{\text{i},3}$) photons emitted from ring 3. A central frequency bin (dashed) is formed by mixing photons from both resonators that consists of a combination of $+1$ sideband from signal (idler) frequency bin  $f_{\text{s},3}$ ($f_{\text{i},3}$) and $-1$ sideband from $f_{\text{s},2}$ ($f_{\text{i},2}$). This creates a single-photon qubit superposition state expressed as:
\begin{equation}
    \ket{\psi}_{\text{s,i}} = \frac{e^{i\phi_{\text{m}}}\ket{1}_{\text{s,i}} + e^{-i\phi_{\text{m}}}\ket{0}_{\text{s,i}}}{\sqrt{2}}
\end{equation}

\thispagestyle{plain}

\indent Using a pair of tunable narrow-band fiber Bragg gratings (FBGs) at the output of the mixing EOM, signal and idler photons at the central bin after mixing are filtered from the rest of the spectrum and routed to the SNSPDs. A coincidence event between the signal and idler channels indicates both photons coming from the same input frequency bins, either $\ket{0}_{\text{s}} \ket{0}_{\text{i}}$ or $\ket{1}_{\text{s}} \ket{1}_{\text{i}}$. Using $\ket{ij}$ to represent $\ket{i}_{\text{s}} \ket{j}_{\text{i}}$, the two-qubit entangled state is written as:
\begin{equation}
    \ket{\psi}_{\text{mix}} = \frac{\ket{00} + e^{i4\phi_{\text{m}}}\ket{11}}{\sqrt{2}}
\end{equation}

\indent The interferogram in Fig.~\ref{Fig6}(\textbf{c}) shows coincidence counts versus modulation phase $\phi_{\text{m}}$ while singles count rates are monitored during the experiment (Fig.~\ref{Fig6}(\textbf{b})). At an average on-chip pump power of 15~\textmu W per resonator, the raw interference visibility without background subtraction $V_{\mathrm{raw}}$ is 92.5\%~$\pm$~3.5\%. Due to our highly efficient sources, the detected coincidence rate at the mixing bin is 23~Hz with a CAR as high as 375, which significantly surpasses previously reported works using single small-FSR resonators operating at higher pump power that deteriorates CAR \cite{imany201850, lu2022bayesian}. 

\thispagestyle{plain}


\indent We proceed to measure CAR and entanglement visibility at higher pump power, and these results are summarized in Fig.~\ref{Fig6}(\textbf{d, e}). As pump power increases, $V_{\text{raw}}$ drops almost quadratically due to increasing background accidentals, eventually dipping below the CHSH threshold. After subtracting background accidentals, the corrected visibility $V_{\text{corr}}$ remains above 90.8\% for all measured power levels. At 250~\textmu W pump power, detected coincidence rate exceeds 7~kHz while $V_{\text{corr}}$ is $92.8\% \pm 0.3\%$, exceeding previous demonstrations despite the higher insertion loss of our measurement setup \cite{clementi2023programmable}.
\\
\indent Finally, to compare our two-photon state to the ideal frequency-bin Bell state, we conduct quantum state tomography by measuring a complete set of 16 projections of the two-qubit states \cite{james2001measurement, takesue2009implementation}. To individually address the phase of the signal and idler input modes, we insert a pulse shaper before the mixing EOM input that separately controls the phase on $\ket{1}_{\text{s}}$ and $\ket{1}_{\text{i}}$. By applying the appropriate phase shift configurations and tuning the pass-bands of fiber Bragg gratings to change the projection basis, we perform a total of 36 coincidence measurements that are used to reconstruct a physical density matrix $\hat{\rho}$ of the quantum state with a maximum likelihood estimation method (Appendix~\ref{Appendix C}). Real and imaginary parts of the estimated density matrix are shown in Fig.~\ref{Fig6}(\textbf{f}) and (\textbf{g}), respectively. For a maximally entangled Bell state $\ket{\Phi^{+}}=\frac{\ket{00}+\ket{11}}{\sqrt{2}}$, the density matrix exhibits fidelity, defined as $\mathcal{F}=\bra{\Phi^{+}} \hat{\rho} \ket{\Phi^{+}}$, exceeding 87\% and 90\% after background correction.
\renewcommand{\arraystretch}{1.4}
\begin{table*}[t]
    \small
    \centering
    \vspace{-10pt}
    \begin{tabular}{cccccc}
        \hline \hline
        \hspace{10pt}\textbf{Ring}\hspace{10pt} & \hspace{10pt}\textbf{Quality factor}\hspace{10pt} & \hspace{10pt}$\boldsymbol{R}_{\textbf{PG}}$~\textbf{(GHz)}\hspace{10pt} & \hspace{10pt}\textbf{CAR}\hspace{10pt} & \hspace{10pt}\textbf{Time-bin visibility}\hspace{10pt} & \hspace{10pt}$\boldsymbol{g}^{\textbf{(2)}}_{\textbf{h}}\textbf{(0)}$\hspace{10pt}\\[-4pt]
        & \hspace{10pt}\textbf{(}$\boldsymbol{Q}_{\textbf{L}}$\textbf{)}\hspace{10pt} & \hspace{10pt}\textbf{@~1~mW}\hspace{10pt} & \hspace{10pt}\textbf{@~10~kHz}\hspace{10pt} & \hspace{10pt}$\boldsymbol{V}_{\textbf{raw}}$ \textbf{@~0.6~MHz}\hspace{10pt} & \hspace{10pt}\textbf{@~0.6~MHz}\hspace{10pt} \\
        \hline
        1&$2.0 \times 10^{5}$ & 0.43 & 4958 & 95.0\%~$\pm$~1.2\% & 0.010~$\pm$~0.002 \\
        2&$3.1 \times 10^{5}$ & 1.01 & 4582 & 94.3\%~$\pm$~1.2\% & 0.023~$\pm$~0.002 \\
        3&$3.0 \times 10^{5}$ & 0.77 & 5563 & 93.4\%~$\pm$~2.0\% & 0.024~$\pm$~0.005 \\
        4&$6.4 \times 10^{5}$ & 2.60 & 846 & 89.0\%~$\pm$~2.0\% & 0.017~$\pm$~0.004 \\
        5&$2.9 \times 10^{5}$ & 0.27 & 1540 & 90.5\%~$\pm$~2.6\% & 0.018~$\pm$~0.005 \\
        \hline  \hline
    \end{tabular}
    \caption{\small \label{Table 1} Summary of device performances for all five resonators. $Q_{\text{L}}$ is the loaded quality factor. Pair generation rate $R_{\text{PG}}$ is normalized to pump power of 1~mW. CAR is measured at approximately 10~kHz internal PGR. Time-bin entanglement visibility $V_{\mathrm{raw}}$ and heralded second-order correlation function $g^{(2)}_{\mathrm{h}}(0)$ are measured at 0.6~MHz internal PGR.
    }
    \vspace{-10pt}
\end{table*}
\newpage

\thispagestyle{plain}

\section{Conclusion and Outlook}
\indent The majority of approaches to improve the brightness of entangled photon-pair sources primarily focus on reducing the losses to enhance the microresonator quality factor \cite{ramelow2015silicon, chen2024ultralow}, reducing the radius to achieve a smaller cavity mode volume \cite{chopin2023ultra}, and utilizing materials with higher nonlinearities \cite{zhao2022ingap, steiner2021ultrabright}. Here, we implement an approach to achieve high entangled-photon pair generation rates by interleaving the quantum combs via multiplexing multiple small-radius microresonators in an AlGaAs platform, leveraging its high nonlinearity and dense integration. Each resonator has an average on-chip pair generation rate efficiency of $\sim1$ GHz/mW$^2$ and independent thermo-optic tuning enables arbitrary mode spacing with sub-GHz precision. To achieve the same level of quantum information spectral density from a single microresonator, radii $\geq$ 50 times larger than those studied here are required, which would reduce the on-chip pair generation rate by 2500-fold. The high quality of the generated quantum states from the microresonator array is verified with up to $95\%$ time-bin entanglement visibility, $>92\%$ ($90\%$) frequency-bin entanglement visibility (Bell-state fidelity), coincidence-to-accidental ratio up to 5,000, and heralded single-photon purity up to $99\%$. 
\\
\indent We envision several near-term applications leveraging the high rate and high capacity of microresonator arrays. As shown in Fig.~\ref{Fig1}(\textbf{a}), peer-to-peer and multi-user entanglement distribution for quantum communications and networking can be enhanced using spectrally multiplexed sources. Previous work using a single AlGaAsOI resonator illustrates the broad spectrum spanning $> 40$~THz with more than 20 sets of time–energy entangled modes and entanglement-based secret key rates up to 8~kbps without multiplexing \cite{steiner2023continuous}. Using a single integrated device with 20 microresonators, an estimated $> 1$~Mbps entanglement distribution rate is possible with DWDM filters at the receiver. Multi-user networks using commercial pulse shapers can act as programmable switches to deliver time-energy entanglement between many unique user nodes in a large-scale quantum network with projected user-to-user rate up to 100~kbps. By integrating electro-optic modulators and phase shifters with the source, frequency-bin encoding can be employed for high-rate secure quantum communications \cite{tagliavacche2024}, which is compatible with existing fiber optic infrastructure and telecommunications components \cite{lu2023frequency}. By utilizing $N$ microresonators, higher-dimensional encoding can be utilized, such as $N$-dimensional qudit states, as well as multi-photon entanglement with $M$ photons across $N$ frequency modes. A major goal moving forward is to reduce the chip-to-fiber coupling loss and the insertion loss of our frequency bin setup (resulting in $\sim20$~dB total loss per channel), primarily due to the off-the-shelf telecommunications components connected through optical fibers. For example, on-chip electro-optic modulators with $\sim 1$~dB \cite{fan2016integrated} and on-chip pulse shapers with $\sim 6$~dB insertion loss and sub-3~GHz resolution \cite{wu2024onchippulseshapingentangled} have been demonstrated. Combining these with our microresonator array and previously demonstrated sub-3-dB chip-to-fiber coupling losses in AlGaAsOI \cite{castro2022expanding} would boost the frequency bin two-qubit entanglement rates to $\sim 0.7$~MHz off-chip and $> 40$~MHz on-chip. Lastly, by using a bi-directional pumping scheme with two bus waveguides, frequency-polarization hyperentanglement is possible by utilizing off-chip polarization control or on-chip polarization splitter-rotators \cite{miloshevsky2024cmos}. Such states could boost the rates and security of various quantum communication protocols, including dense coding and entanglement distillation \cite{barreiro2008beating,simon2002polarization}, as well as serve as a resource for cluster state generation for loss-tolerant quantum networking and computing.
\section*{Funding}
\indent This work was supported by AFOSR (Grant No. FA9550-23-1-0525), the NSF Quantum Foundry (Grant No. DMR-1906325), the NSF CAREER Program (Grant No. 2045246), and the Eddleman Center for Quantum Innovation. L.T. acknowledges support from the NSF Graduate Research Fellowship Program. M. L. acknowledges PNRR MUR project "National Quantum Science and Technology Institute" -- NQSTI (Grant No. PE0000023).
\section*{Acknowledgment}
\indent A portion of this work was performed in the UCSB Nanofabrication Facility, an open access laboratory.
\section*{Disclosures}
\indent The authors declare no conflicts of interest.
\section*{Data Availability}
\indent The data that support the figures in this paper and other findings of this study are available from the corresponding author on reasonable request.
\titleformat{\section}{\bfseries\small\centering}{Appendix \thesection.}{10pt}{}
\begin{appendices}

\thispagestyle{plain}

\section{Coincidence Data Fitting}\label{Appendix A}
\begin{figure}[h]
    \centering
    \includegraphics[width=0.83\linewidth]{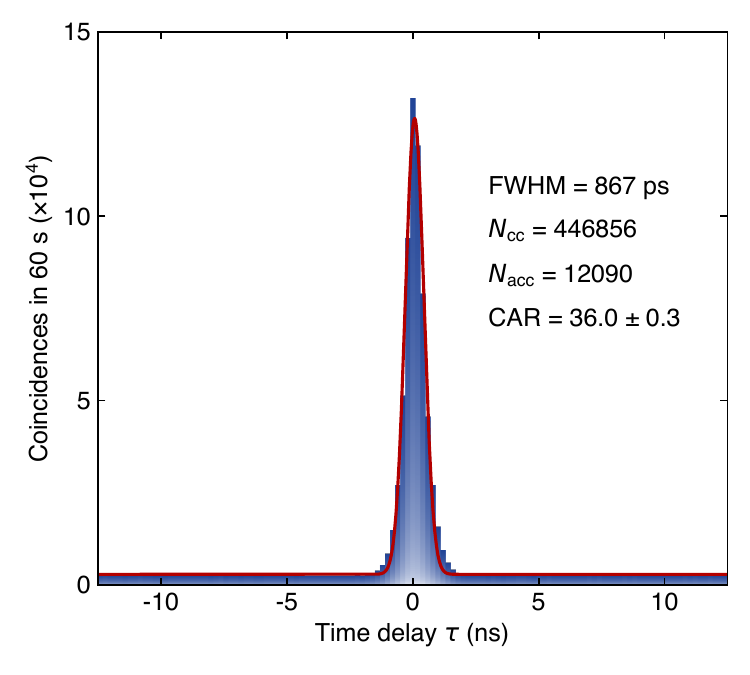}
    \vspace{-15pt}
    \caption{\small \label{Fig7} A representative coincidence histogram showing the coincidence counts as a function of time delay $\tau$ over an integration time of 60 seconds. Red line is a Gaussian fit to the experiment data. 
    }
    \vspace{-10pt}
\end{figure}

\begin{figure}[hbt]
    \centering
    \includegraphics[width=0.83\linewidth]{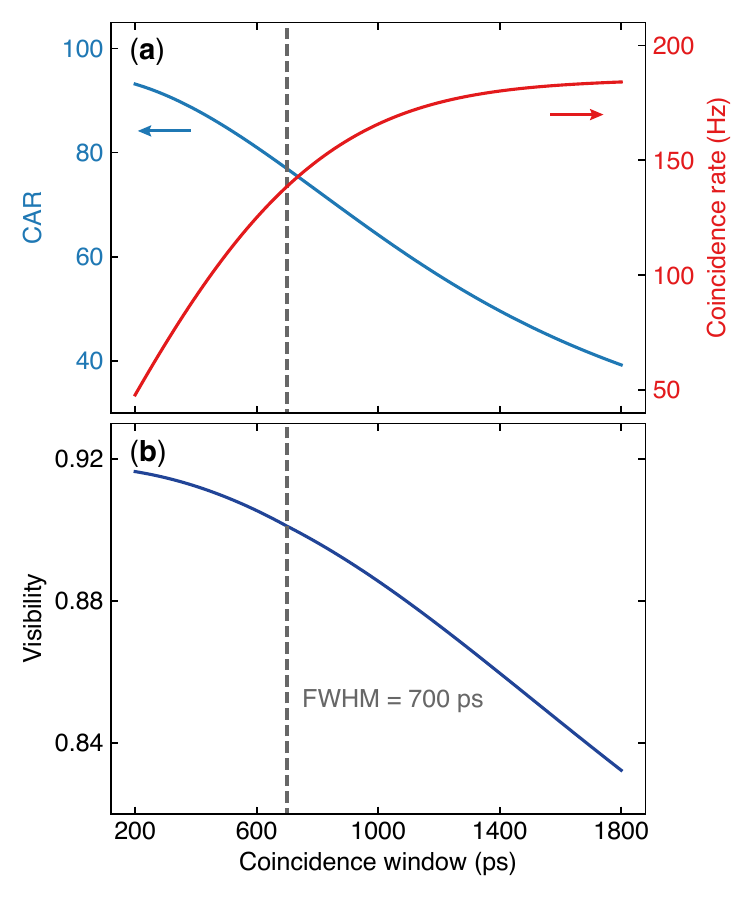}
    \vspace{-15pt}
    \caption{\small \label{Fig8} Effect of varying coincidence window on the calculated (\textbf{a}) CAR and detected coincidence rate, and (\textbf{b}) visibility for the frequency-bin entanglement experiment.
    }
    \vspace{-10pt}
\end{figure}

\indent A representative coincidence histogram measured at an on-chip pump power of 67.3~\textmu W (ring 2) is shown in Fig.~\ref{Fig7}. The coincidence counts are integrated for 60 seconds, and longer integration times of up to 60 minutes are implemented at lower power to reach non-zero accidental counts. An electronic delay is applied on our TCSPC to compensate for the length difference between signal and idler channels. We use a Gaussian function to fit the coincidence data, and the FWHM of coincidence peak is calculated using the fitted standard deviation $\sigma$ of the Gaussian distribution via FWHM = $(2\sqrt{2\ln{2}})\sigma$. The FWHM reflects the single-photon coherence time, which is inversely proportional to the linewidth of the cavity resonances. Here, using the FWHM of 867~ps as the two-photon coincidence time window, we calculate $N_{\text{cc}} = 446856$, $N_{\text{acc}} = 12090$, and $\text{CAR} = 36.0 \pm 0.3$.
\vspace{-10pt}
\\
\indent The choice of the time window has a non-trivial impact on coincidence counting and the CAR. Figure~\ref{Fig8} shows the effect of varying the coincidence window on CAR and detected coincidence rate (Fig.~\ref{Fig8}(\textbf{a})), as well as raw entanglement visibility (Fig.~\ref{Fig8}(\textbf{b})) for one set of data taken from the frequency-bin entanglement experiment (on-chip pump power of 33.7~\textmu W). Coincidence rate and CAR shown here (as well as in Fig.~\ref{Fig6}(\textbf{d})) are taken at the modulation phase $\phi_{\text{m}}$ with maximum coincidence counts (constructive interference). Although the coincidence rate increases with the coincidence window width, CAR and visibility degrade for windows larger than the FWHM, which is due to more accidental counts contributing to the coincidences. We choose the FWHM of 700~ps as the coincidence window, yielding the coincidence rate reported for this set of data equal to $138.7$~Hz with a CAR of 76.8, and raw visibility of 90.1\%.
\vspace{-10pt}
\section{Experiment Setup}\label{Appendix B}
\begin{figure*}[htb]
    \centering
    \includegraphics[width=\textwidth]{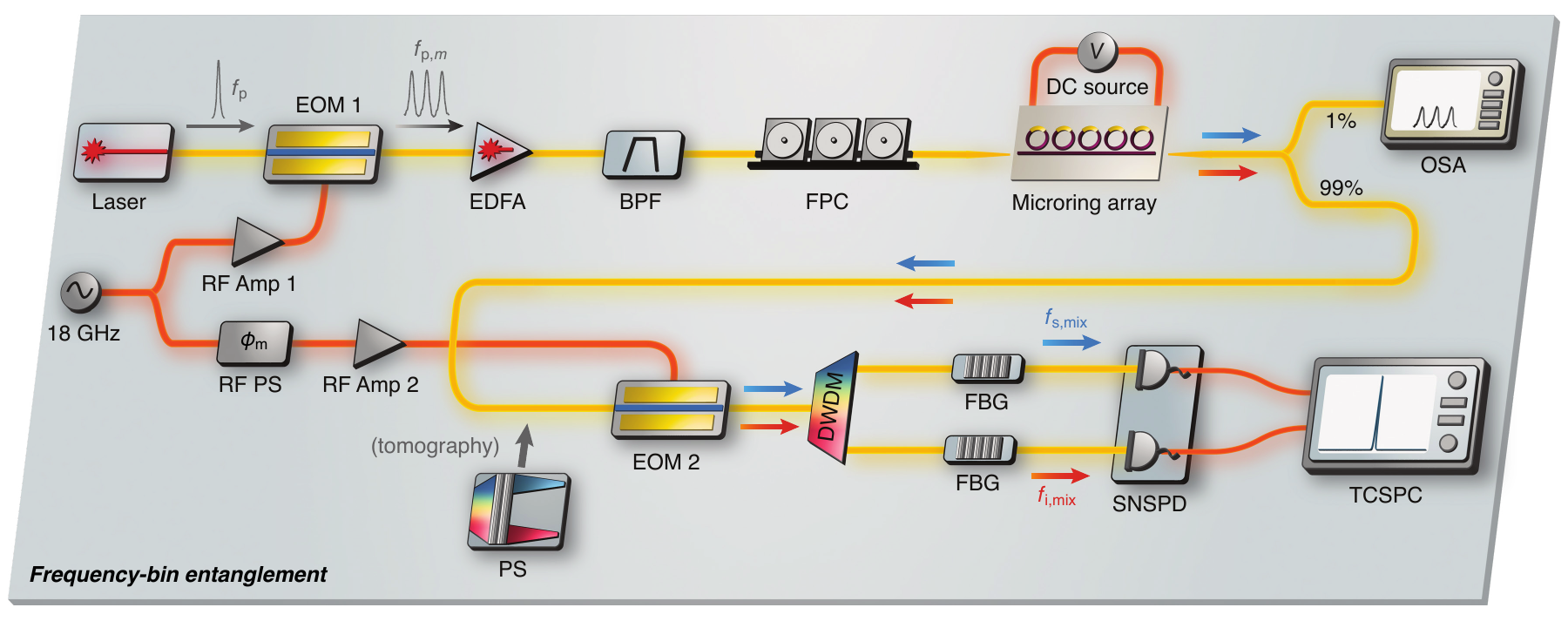}
    \vspace{-15pt}
    \caption{\small \label{Fig9} A schematic illustration of the experiment setup for frequency-bin entanglement. A pulse shaper is inserted before EOM 2 for tomography measurements. (RF PS, RF phase shifter; FBG, fiber Bragg grating filter.)
    }
    \vspace{-10pt}
\end{figure*}
\indent Figure~\ref{Fig9} shows the experiment setup used for frequency-bin entanglement and quantum state tomography. Pump light ($f_{\text{p}}$) from a tunable continuous wave laser (Toptica CTL) is modulated using a commercial 20~GHz lithium niobate phase modulator (EOM~1, EOSpace) driven by an 18~GHz RF signal generator (Keysight PSG) amplified with a low-noise RF amplifier (RF Amp~1, RF Lambda). An erbium-doped fiber amplifier (EDFA, Amonics) optically amplifies the 18~GHz EO comb ($f_{\text{p}, m}$) generated from phase modulation. Laser amplified spontaneous emission is spectrally filtered with a series of three tunable bandpass filters (BPF, Agiltron) and coupled on-chip with a lensed fiber. A fiber polarization controller (FPC) aligns pump polarization to the transverse electric (TE) waveguide mode. Resonances of the two resonators are controlled by a multichannel DC current source (Qontrol) and pumped with the $\pm 1$ sidebands of the EO comb. While 1\% of the light containing residual pump is monitored with an optical spectrum analyzer (OSA, Yokogawa), 99\% of the output is sent to a second phase modulator (EOM~2) where the frequency mixing of generated photon pairs takes place. EOM~2 is driven by the same 18~GHz RF signal with $\phi_{\text{m}}$ controlled by an RF phase shifter (RF Lambda). Photon pairs after mixing are separated by DWDM filters (Fiberdyne Labs) into signal and idler channels, which are then filtered with a pair of FBGs (Advanced Optics Solutions). The mixed signal ($f_{\text{s,mix}}$) and idler ($f_{\text{i,mix}}$) photons are routed to a pair of SNSPDs (PhotonSpot), and coincidence events are time-tagged with a TCSPC (Swabian Instruments). For quantum state tomography, the same pulse shaper (Finisar Waveshaper) used in JSI measurements is inserted at the input of EOM~2 to individually control the phase of the input states.

\thispagestyle{plain}

\section{Quantum State Tomography}\label{Appendix C}
\indent Quantum state tomography of the generated quantum state is performed using the pulse shaper to apply individual phase shifts on the signal and idler photons ($\ket{1}_{\text{s}}$ and $\ket{1}_{\text{i}}$, respectively) generated from ring 3 before entering the mixing EOM. The RF phase shifter is kept at a constant phase shift throughout the measurement. Similar to the case for time-bin entanglement as discussed in \cite{takesue2009implementation}, we consider four possible single-photon states $\ket{0}$, $\ket{1}$, $\ket{+}$, $\ket{\text{L}}$, where $\ket{+}=\frac{\ket{0} + \ket{1}}{\sqrt{2}}$ and $\ket{\text{L}}=\frac{\ket{0} + i\ket{1}}{\sqrt{2}}$. Since these states correspond to different frequency bins, the projection measurement is done by spectrally filtering the chosen bin with the tunable FBG filters. Photon detection at the highest (lowest) frequency bin of $+1$ ($-1$) sideband generated from ring 2 (ring 3) corresponds to a projection measurement of state $\ket{0}$ ($\ket{1}$). Alternatively, photon at the central bin is described as a pure state:
\begin{equation}
    \ket{\psi} = \frac{\ket{0}+e^{i\theta}\ket{1}}{\sqrt{2}}
\end{equation}
when the pulse shaper applies phase shift of $\theta$ on the input photon $\ket{1}$. Thus, the projection basis is set to $\ket{+}$ when $\theta = 0$ and to $\ket{\text{L}}$ when $\theta = \pi/2$. Since there are four projection bases per photon, we perform a total of 16 combinations of projection measurements with $\ket{\Psi_{v}}=\ket{A}_{\text{s}}\ket{B}_{\text{i}}$ where $v \in \{1,2,\dots, 16\}$ and $A, B \in \{1, 2, +, \text{L}\}$. While the FBGs in the signal and idler channels are tuned to select the corresponding frequency bins, four types of phase settings are imposed on the input photons with $(\theta_{\text{s}}, \theta_{\text{i}}) \in \{(0, 0),(0, \pi/2),(\pi/2, 0),(\pi/2, \pi/2)\}$ \cite{imany201850}. These phase settings can be expressed in the projection bases $\ket{+}$ and $\ket{\text{L}}$ as $\{\ket{++}, \ket{+\text{L}}, \ket{\text{L}+}, \ket{\text{L}\text{L}}\}$, hence not all the projections $\ket{\Psi}_{v}$ necessitate coincidence measurements in all four phase settings. For example, $\ket{\Psi_{7}}=\ket{+}_{\text{s}}\ket{+}_{\text{i}}$ is projected only when $(\theta_{\text{s}}, \theta_{\text{i}}) = (0,0)$. We integrate coincidence counts for 60 seconds for all 36 permutations of bases and phase configurations, denoted as $n_{v,k}$ with $k \in \{1,2,3,4\}$ for four phase settings. The total coincidence counts for each projection state $\ket{\Psi_{v}}$ is $N_{v}=\sum_{k=1}^{4} n_{v,k}$, and is equivalent to $N_{v}=\mathcal{N}\bra{\Psi_{v}}\hat{\rho}\ket{\Psi_{v}}$, where $\mathcal{N}$ is a constant determined by the PGR and detection efficiencies, and $\hat{\rho}$ is the density matrix describing the quantum state. 
\\
\indent Although $\hat{\rho}$ can be calculated via linear matrix operations, the result is often non-physical due to imperfect experiment data and statistical variations of the photon counting process \cite{james2001measurement}. Therefore, we implement a maximum likelihood estimation method, where a physical density matrix $\hat{\rho}_{\text{p}}$ is introduced and the matrix elements are numerically optimized to maximize its relation to the measured data. We perform a multivariate optimization protocol to minimize the likelihood function:
\begin{equation}
    \mathcal{L} = \sum_{v=1}^{16} \frac{(\mathcal{N}\bra{\Psi_{v}}\hat{\rho}_{\text{p}}\ket{\Psi_{v}} - N_{v})^2}{2\mathcal{N}\bra{\Psi_{v}}\hat{\rho}_{\text{p}}\ket{\Psi_{v}}}
\end{equation}
which equivalently maximizes the probability that the quantum state described by $\hat{\rho}_{\text{p}}$ generates the measured data $n_{v,k}$.
\end{appendices}

\thispagestyle{plain}

\def\bibsection{\section*{References}}
\bibliography{references.bib}

\thispagestyle{plain}

\end{document}